\documentclass[a4paper]{toledoStyle}
\usepackage{epsfig}

\begin{document}

\title{ASTRO-F Survey as Input Catalogues for FIRST}

\author{Takao Nakagawa\inst{1}}

\institute{Institute of Space and Astronautical Science (ISAS),
  3-1-1 Yoshinodai, Sagamihara, Kanagawa 229-8510, Japan}

\maketitle 

\begin{abstract}

ASTRO-F is the second Japanese space mission for infrared astronomy
and is scheduled to be launched into a sun-synchronous polar orbit
by the Japanese M-V rocket in February 2004. 
ASTRO-F has a cooled 67 cm telescope with
two focal plane instruments: one is the Far-Infrared
Surveyor (FIS) and the other is the Infrared Camera (IRC). 

The main purpose of FIS is to perform the all-sky survey with
4 photometric bands in the wavelength range 
of 50 - 200 $\mu$m. The advantages
of the FIS survey over the IRAS survey are (1) higher 
spatial resolution ($30''$ at 50-110 $\mu$m and $50''$ at 110-200 $\mu$m)
and (2) better sensitivity by one to two orders of magnitude. 
The FIS survey will provide
the next generation far-infrared survey catalogs, which will be
ideal inputs for observations by FIRST.

The other instrument, IRC, 
will make deep imaging and low-resolution spectroscopic
observations in the spectral range of $1.8-26 \mu$m.
The IRC will make large-area surveys
with its wide field of view ($10' \times 10'$), and 
will be complementary with the
FIRST observations at longer wavelengths.

\keywords{Galaxies: formation -- Stars: formation -- Planets: formation -- 
Missions: ASTRO-F} 
\end{abstract}

\section{ASTRO-F Mission}

ASTRO-F (Murakami 1998) is the second Japanese space mission for
infrared astronomy, following the first successful mission
IRTS (Infrared Telescope in Space, Murakami et al. 1996).

\begin{figure}[htb]
\begin{center}
\epsfig{file=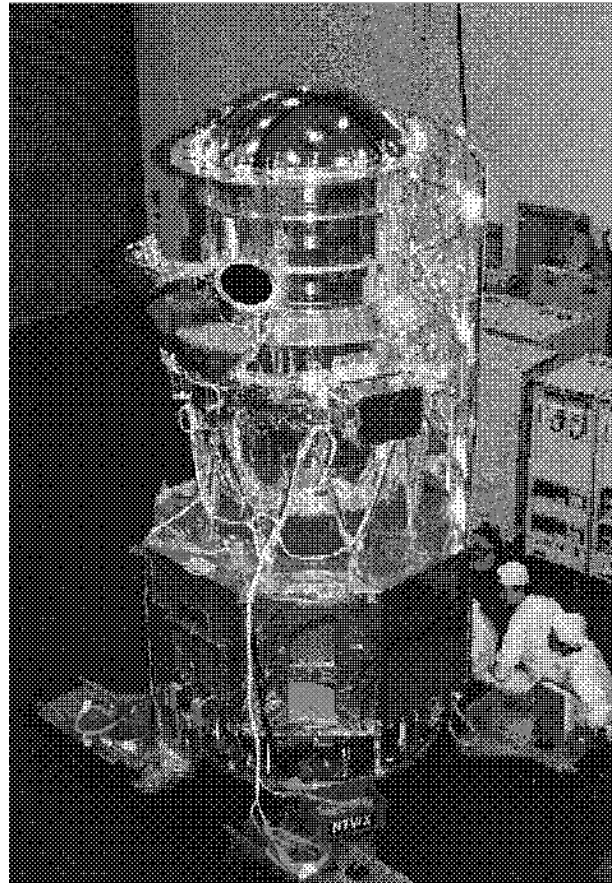, width=8cm}
\end{center}
\caption{Thermal test model of the ASTRO-F}
\label{fig:ttm}
\end{figure}

\begin{figure*}
\begin{center}
\epsfig{file=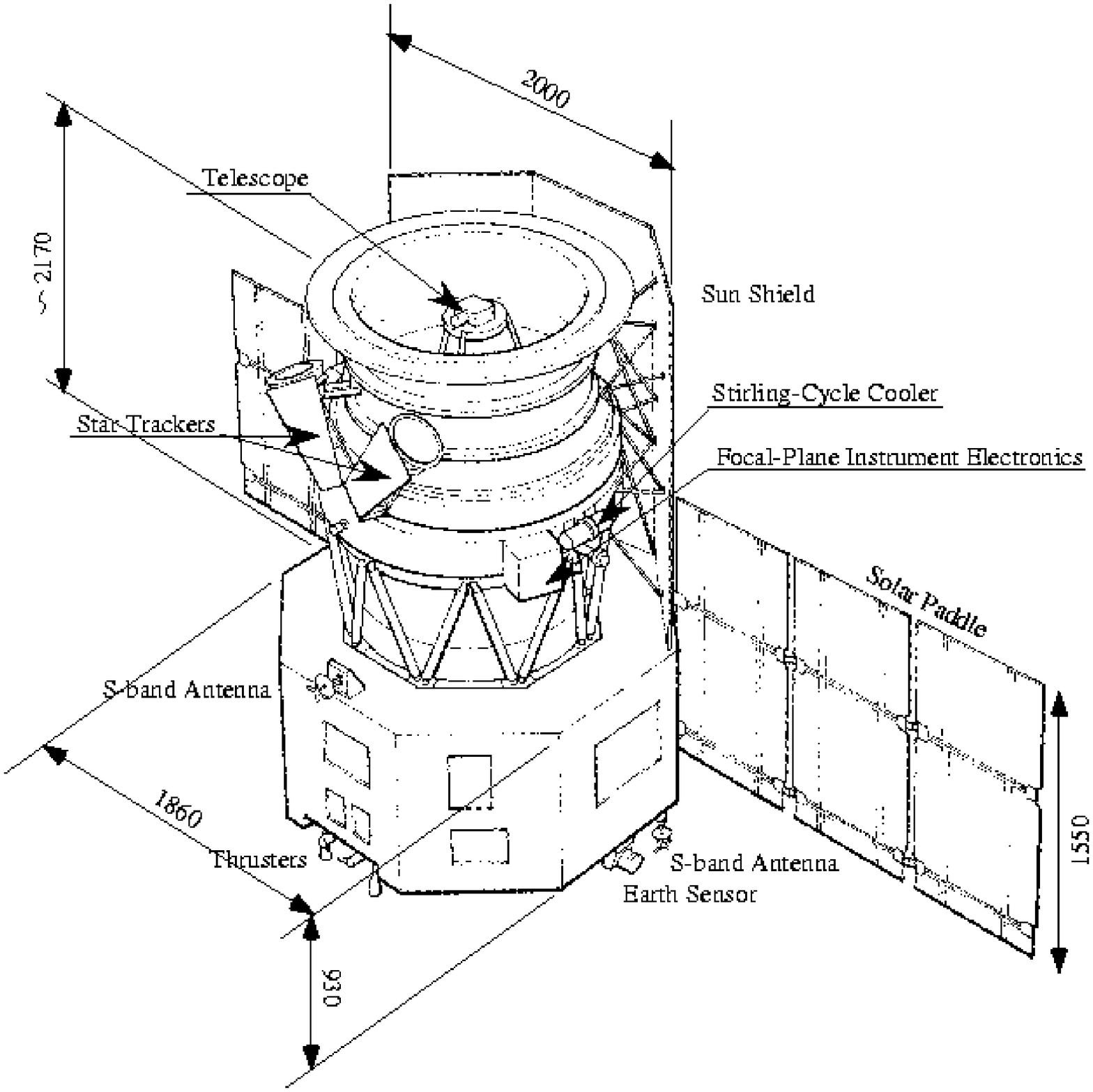, width=13cm}
\end{center}
\caption{The ASTRO-F system in the observation configuration
	after the aperture lid is jettisoned.}
\label{fig:all}
\end{figure*}

The ASTRO-F is designed as the
second-generation survey mission. 
The previous all sky survey by the
Infrared Astronomy Satellite (IRAS) (Neugebauer et al. 1984)
was a great success. Many new phenomena, such as 
infrared galaxies and Vega-like stars,  were found in 
the survey, and the IRAS survey catalogues have been essential
tools in many fields of astronomy.

However, the spatial resolution (a few arcminutes) 
and the sensitivity (about 1 Jy) of the IRAS survey are not
good enough, compared to those of the surveys at other wavelengths.
Hence we plan the next generation infrared sky survey with ASTRO-F
by taking the advantage of the recent development of detector
technology. 

\begin{table}[bt]
\caption{Specifications of ASTRO-F}
\label{tab:spec}
\begin{center}
\begin{tabular}{ll}
\hline
\hline
Effective Apeture & 67~cm \\
		& (Diffraction Limit at 5 $\mu$m)\\
$T_{\rm Mirror}$ & 5.8 K \\
Wavelength Range & $1.8-200 \mu$m \\
Orbit & Sun Synchronous Polar Orbit\\ 
	& above the twilight zone\\
Cryogenics & 170 liters of liquid helium\\
	& Two sets of 2-stage Stirling Cycle\\
Total Mass & 960 kg (wet) \\
Launch Vehicle & M-V Rocket\\
Launch Year & early 2004\\
\hline
\end{tabular}
\end{center}
\end{table}

ASTRO-F has a  67 cm telescope cooled to 5.8~K,
and covers wide wavelength range from K-band to 200 $\mu$m
with two focal plane instruments: Far-Infrared
Surveyor (FIS) and Infrared Camera (IRC). 
FIS will perform the
all-sky survey with 4 photometric bands at the wavelength range of 
50 -- 200 $\mu$m using two-dimensional Ge:Ga detector arrays. 
The sensitivity and the spatial resolution 
of the ASTRO-F/FIS all sky survey are much better than those of
the IRAS survey. 
IRC is for the near- and mid-infrared ranges,
and large-format arrays are employed for deep sky surveys in selected
sky regions. 

The scientific targets of the ASTRO-F range from distant galaxies to
near-by objects in the solar system. 
The results of the new infrared survey by ASTRO-F will serve as
valuable input catalogues for large-aperture space telescopes
such as FIRST (Pilbratt 2000) and also for 8--10~m class
ground-based telescopes.

Figure~1 shows a picture of the thermal test model of
the ASTRO-F satellite, Figure~2 shows a diagram
of ASTRO-F in the observation configuration, 
and Table~1 shows the specifications of ASTRO-F

\section{ASTRO-F Hardwares}

\begin{figure}[htb]
\begin{center}
\epsfig{file=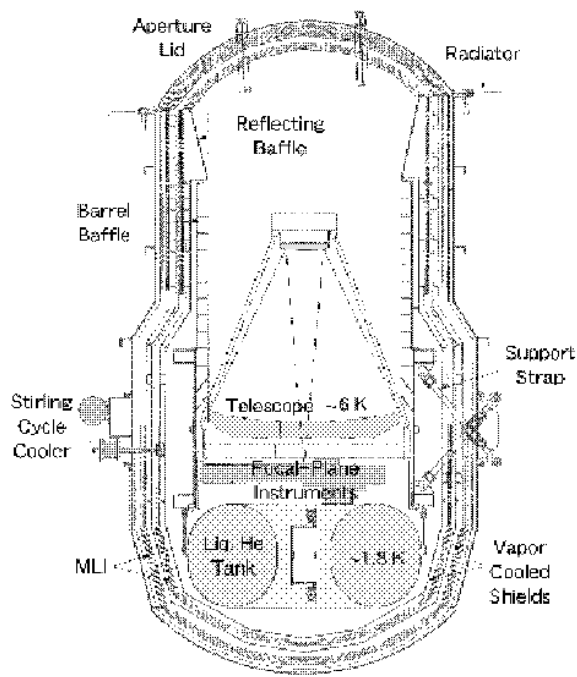, width=8cm}
\end{center}
\caption{Cross-cut view of the ASTRO-F cryostat, which is a hybrid system
with relatively small amount (170 l) of liquid Helium and two mechanical
cryocoolers}
\label{fig:cross}
\end{figure}

\subsection{Hybrid Cryogenics}

To reduce the total weight of the cryostat,
we employed a hybrid cryogenics design; the telescope
and focal plane instruments are to be cooled by liquid helium
with the help of mechanical cryocoolers.

Figure~3 shows a cross-sectional view of the ASTRO-F cryostat 
with the 67 cm telescope installed.
The outer shell of the cryostat is thermally isolated from
the spacecraft and is cooled below 200 K by radiation. The
cryostat has two vapor-cooled shields (VCS). The inner VCS is
cooled not only by evaporated helium gas but also 
by two sets of 2-stage Stirling-Cycle coolers.
This additional cooling power provided by the coolers
increases the life time of the liquid helium by a factor of two.
Although the amount of liquid helium is small (170 liter), 
the expected hold time of liquid helium in orbit is as long as 550
days due to the cryocoolers.
One more advantage of using mechanical coolers is that,
as long as the mechanical coolers work properly, 
we can continue observations at least in the near-infrared
even after liquid helium runs out.
The cryocoolers are now under extensive tests, and
the expected life time of the coolers is
longer than 2 years.

The telescope and most of the focal-plane instruments
are cooled by the evaporated helium gas, and the expected temperature
is 5.8 K. 
Far-infrared detector arrays require lower temperatures, and are 
cooled to 1.8~K by the direct thermal connection to the helium tank.

\subsection{Light-weight Telescope}

\begin{figure}[htb]
\begin{center}
\epsfig{file=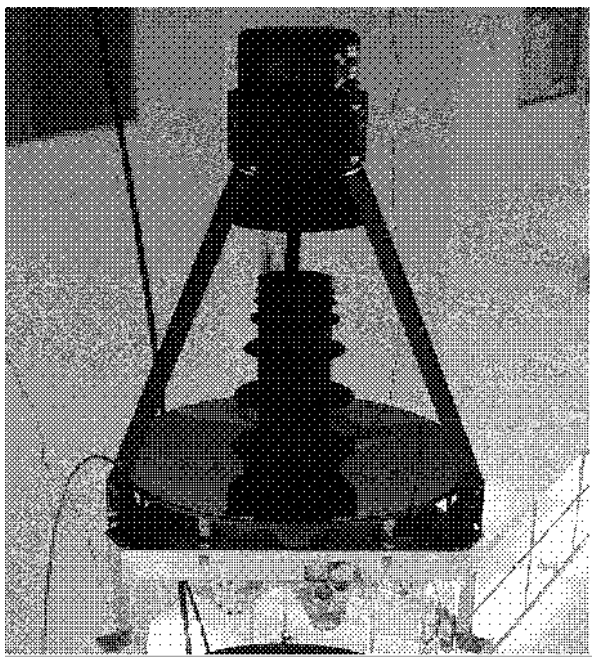, angle=270, width=7.8cm}
\end{center}
\caption{The telescope assembly of ASTRO-F. Both the primary and secondary
	mirrors are made of SiC.}
\label{fig:telescope}
\end{figure}

ASTRO-F telescope is a Ritchey-Chretien system with a 
effective aperture size of 67 cm. Figure~4 shows
the whole telescope assembly, which is to be cooled down 
to 5.8~K.
The goal of the image quality is diffraction-limited
performance at the wavelength of 5 $\mu$m, including the aberration of
the camera optics at the focal plane. 

In order to reduce the total weight of the telescope system,
we have employed silicon carbide (SiC) for mirror material
because of its large Young's modulus and high thermal
conductivity. 
Both the primary and secondary mirrors have a sandwich 
structure which consists of  
porous SiC (3 mm thick) as a core 
and CVD coat of SiC (0.5 mm thick) on surfaces.
Porous SiC can be easily machined for
light-weight structures, and
CVD SiC coat is dense and strong enough
for smooth polishing. 
This structure reduces the weight of the primary mirror
to 11~kg. 
The mirrors are coated with Au for good reflectivity
with ZnS overcoat for protection.

We plan to measure  the surface figure of the primary mirror
and the whole telescope assembly both at ambient temperature
and at liquid helium temperature.
Please see Onaka, Sugiyama \& Miura (1998) and 
Kaneda, Onaka \& Yamashiro (2000) for
details of the ASTRO-F telescope and its developing program.

\subsection{Focal Plane Instruments}

\begin{figure*}[tbp]
\begin{center}
\epsfig{file=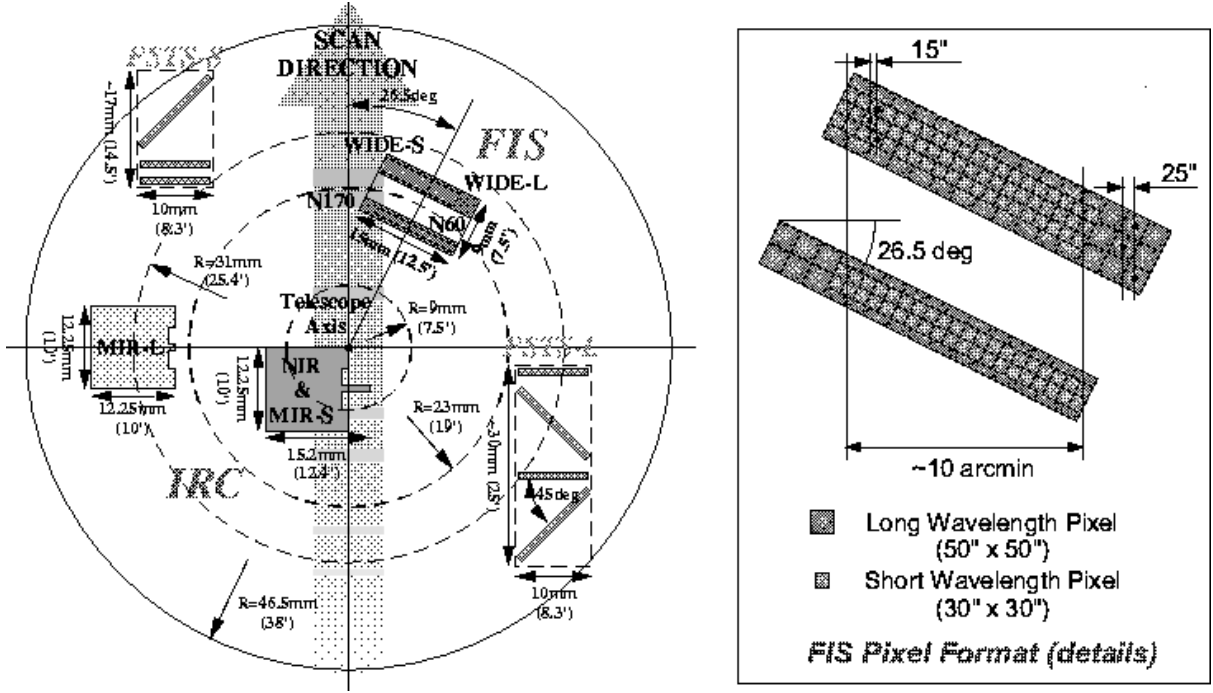, width=16cm} 
\end{center}
\caption{The configuration of the focal plane (left). The vertical
	arrow shows the scan direction. IRC consists
	of three channels, and the NIR and 
	the MIR-S channels share the same entrance aperture. 
	FIS has two detector modules, and they also share the same
	entrance aperture. Two sets of FSTS (Focal Plane Star Sensor)
	are used for pointing reconstruction during the survey mode.
	The detector arrays of FIS are tilted against the scan direction
	by 26.5$\degr$ (right) to improve the sampling frequency
	in the cross-scan direction.}
\label{fig:fpi}
\end{figure*}

\begin{figure*}[htb]
\begin{center}
\epsfig{file=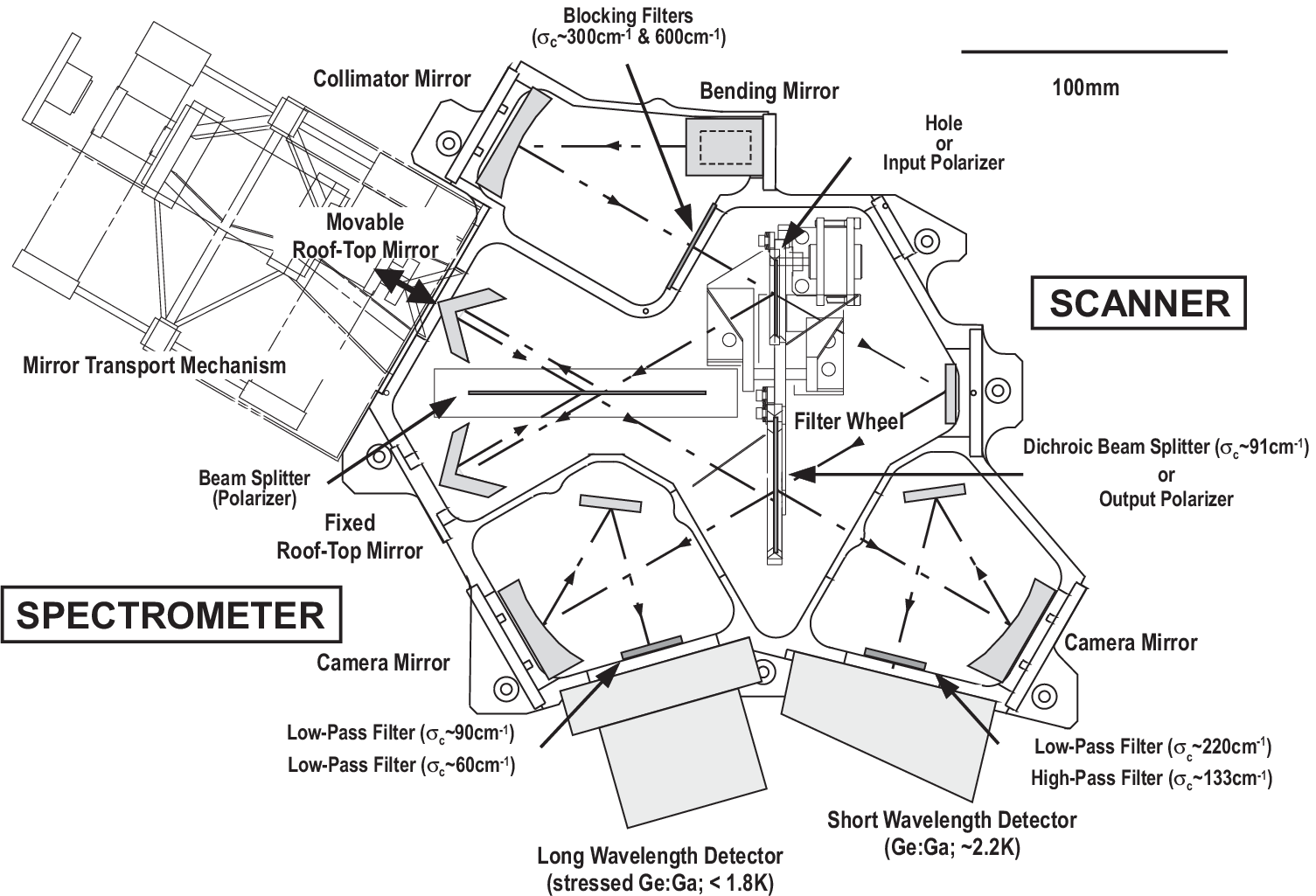, width=15cm}
\end{center}
\caption{Optical layout of FIS. The entrance aperture is
	below the bending mirror.
	FIS has two modes: scanner (or photometric imager) 
	and (imaging) spectrometer. We can switch the modes
	by changing the filter wheel position.}
\label{fig:fis}
\end{figure*}

ASTRO-F has two focal instruments: Far-Infrared
Surveyor (FIS) for far-infrared and 
Infrared Camera (IRC) for near- and mid-infrared.

These two instruments share the focal plane 
as shown in the left figure of Figure~5.
The vertical
arrow shows the scan direction. IRC consists
of three channels, but the NIR and 
the MIR-S channels share the same entrance aperture,
and there are only two apertures in total for IRC. 
FIS has two detector modules, and they also share the same
entrance aperture. Two sets of FSTS (Focal Plane Star Sensor)
are used for pointing reconstruction during the survey mode.
The goal of the pointing reconstruction
is to determine the
telescope direction with the uncertainty less that 
$3''$, which is required for the identification 
of observed infrared sources at other wavelengths.

\subsection{FIS: Far-Infrared Surveyor}

FIS (Kawada 2000, Takahashi et al. 2000) is designed
primarily to perform the all-sky survey in 4 photometric bands at
wavelength of 50 - 200 $\mu$m.  FIS also has spectroscopic
capability as a Fourier-Transform spectrometer.

Figure~6 shows 
the optical layout of FIS.
The incident beam from the telescope comes
from the entrance aperture below the bending mirror.
The incident beam is paralleled by the collimator
mirror. This parallel beam goes to the filter wheel. 
This filter
wheel determines the mode of operation, i.e. scanner (=photometric 
imager) or spectrometer.

In the scanner mode, the incident filter on the filter
wheel is just a hole.  The parallel beam
goes through the hole and reflected by the flat
mirror.   Then the beam is divided into two in the spectral
domain by a dichroic beam splitter on the same filter wheel (but at
the opposite side).  The radiation longer
than 110 $\mu$m goes through the filter and is
concentrated onto the long wavelength detector module.
The shorter wavelength radiation is reflected by the dichroic beam
splitter and is focused on the short wavelength detectors module. 
Each detector module consists of two detector arrays.
Another set of filters just
in front of the detector arrays determines
the effective photometric bands.   
Table~2 lists the four photometric bands.
The optical efficiency of each band is about 40 $\%$.   

\begin{table} [hbt]    
\begin{center}        
\caption{Specifications of FIS}
\label{tab:fis_spec} 
\begin{tabular}{cccc}  
\hline 
\hline 
\rule[-1ex]{0pt}{3.5ex}  Band & Wavelength & Pixel FOV & Array\\[-5pt] 
\rule[-1ex]{0pt}{3.5ex}  & ($\mu$m) & (arcsec)&  \\ 
\hline 
\rule[-1ex]{0pt}{3.5ex}  N60 & $50-75$ &
30 & $20 \times 2$ \\
\rule[-1ex]{0pt}{3.5ex}  Wide-S & $50-110$ &
30 & $20 \times 3$ \\
\rule[-1ex]{0pt}{3.5ex}  Wide-L & $110-200$ &
50 & $15 \times 3$ \\
\rule[-1ex]{0pt}{3.5ex}  N170 & $150-200$ &
50 & $15 \times 2$ \\
\hline 
\end{tabular} 
\end{center}
\end{table}

For the spectrometer mode of FIS,
we use a polarized Michelson interferometer
(Martin-Puplett type).
We need three polarizing filters: input, beamsplitter, 
and output.
The input and output polarizing filters are on the
filter wheel and are to be selected for the spectrometer mode.   
In this mode, the parallel beam is 
reflected by the input polarizing filter.
Then the polarized beam goes into the interferometer,
and are divided by the polarizing beam splitter whose polarizing direction is
rotated 45$^{\circ}$ against the polarity of the incident beam. 
Two beams divided by the beam splitter are reflected by two sets
of roof-top mirrors. We move one set of the mirrors
to change the optical path difference between the two beams.
The two beams are combined again on the beam splitter, and the
combined beam goes onto the output polarizing filter.   
By the output polarizing filter, the beams are divided in the
polarity domain, and
concentrated by the camera mirrors on each detector module.   
The maximum optical
path difference is about 50 mm, which corresponds to a spectral
resolution of 0.2 cm$^{-1}$.
The big advantage of the spectrometer mode of FIS is that
it is an imaging spectrometer suitable for spectroscopic
mapping observations of extended sources.

Each detector module consists of two bands, and
FIS have four bands in total as is shown in Table~2.
The detector arrays for WIDE-S
and N60 are Ge:Ga detector array which is connected to the 
cryogenic readout electronics (Hirao et al. 2001)
directly by indium (Hiromoto et al. 1998).
For the detector arrays of WIDE-L and N170, we use
compact stressed Ge:Ga detector arrays (Doi et al. 2000).

The pixel size of each detector array is
comparable with the size of the diffraction pattern of the 
main mirror at each band.
Hence if we scan the sky with the detector array whose
minor axis is 
parallel to the scan direction, the sampling frequency in the
cross-scan direction is
below the Nyquist sampling frequency. In other words,
the spatial resolution in the cross-scan direction is
worse than that of the diffraction limited size.
Hence, we tilted the detector array by 26.5$^{\circ}$ as shown
in the right figure of Figure~5, 
to increase the sampling frequency and thereby to improve the
spatial resolution in the cross-scan direction.
This tilt guarantees the Nyquist sampling
for each scanning observations, which is critical
requirement to achieve the diffraction limited spatial resolution.

\begin{table} [htb]    
\begin{center}        
\caption{Specifications of IRC}
\label{tab:irc_spec} 
\begin{tabular}{cccc}  
\hline 
\hline 
\rule[-1ex]{0pt}{3.5ex}  Channel & Wavelength & Pixel FOV & Array\\[-5pt] 
\rule[-1ex]{0pt}{3.5ex}  & ($\mu$m) & (arcsec)&  \\ 
\hline 
\rule[-1ex]{0pt}{3.5ex}  NIR & $1.8-5$ &
1.46 & $512 \times 412$ InSb \\
\rule[-1ex]{0pt}{3.5ex}  MIR-S & $5-12$ &
2.34 & $256 \times 256$ Si:As \\ 
\rule[-1ex]{0pt}{3.5ex}  MIR-L & $12-26$ &
2.34 & $256 \times 256$ Si:As \\ 
\hline 
\end{tabular} 
\end{center}
\end{table}

\subsection{IRC: InfraRed Camera}

\begin{figure*}
\begin{center}
\epsfig{file=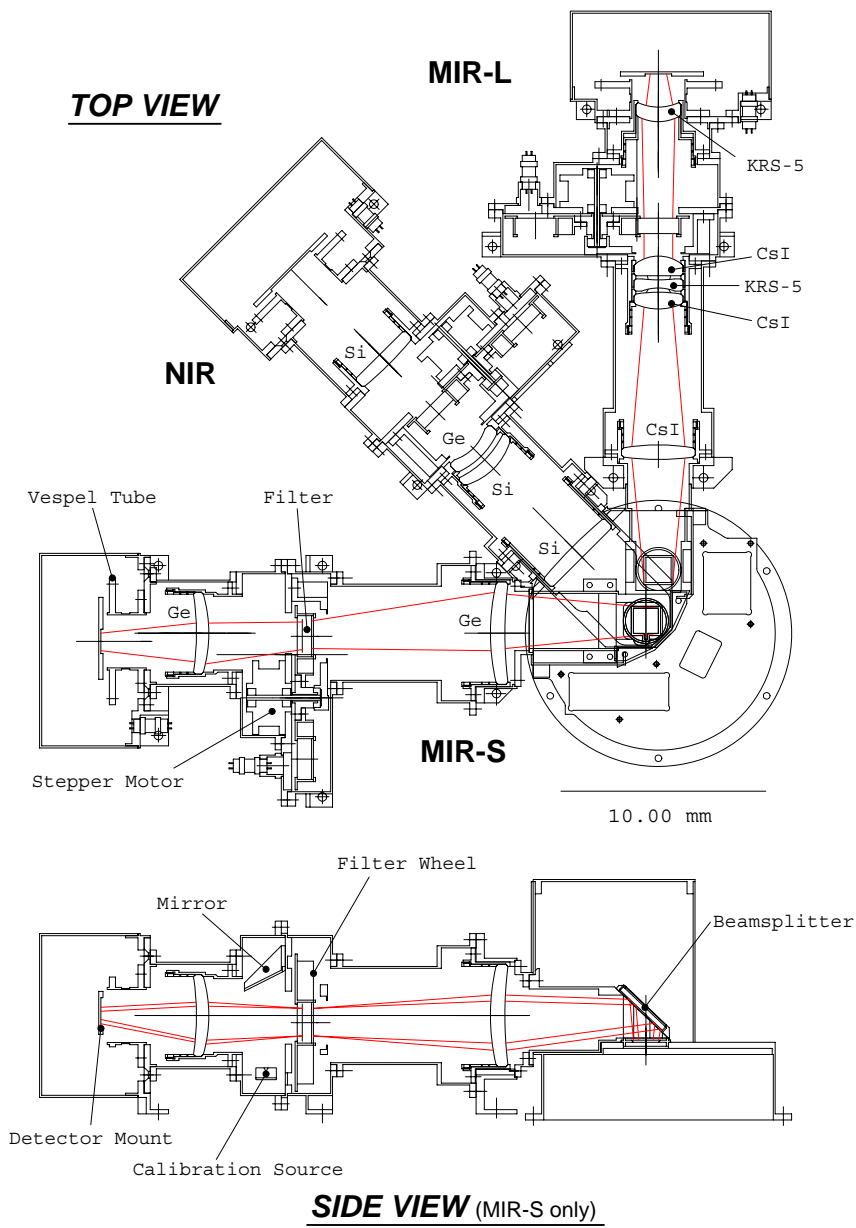, width=13cm}
\end{center}
\caption{Top view of IRC (top), and side view of MIR-S (bottom), one 
	of the three channels of IRC. The optical configuration
	for each channel is shown together with the lens material.}
\label{fig:irc}
\end{figure*}

IRC (Watarai et al. 2000, Onaka et al. 2000) 
is another focal-plane instrument onboard ASTRO-F.  
It is designed for deep imaging and low-resolution  
spectroscopic observations 
in the near- to mid-infrared from $1.8 - 26 \mu$m.

The big advantage of ASTRO-F/IRC
is that it can make deep photometric and spectroscopic 
surveys with wide field-of-views.
It will provide a important database for many fields
of astronomy.

Figure~7 shows the whole
structure of IRC.
IRC consists of three channels: NIR, MIR-S, and MIR-L.
Table~3 shows the details of each channel.
Each channel has
a large-format detector array (Table~3),
which enables nearly diffraction-limited spatial resolution 
at each channel with a wide field-of-view of $10' \times 10'$.
The NIR and MIR-S channels observe the same sky
as is shown in Figure~5,  
while the MIR-L observes the sky about $20'$ away from the 
NIR/MIR-S position. 

\section{ASTRO-F Observations}

\subsection{Modes of Observations}

\begin{figure}[htb]
\begin{center}
\resizebox{8cm}{!}{\includegraphics*[32,419][515,819]{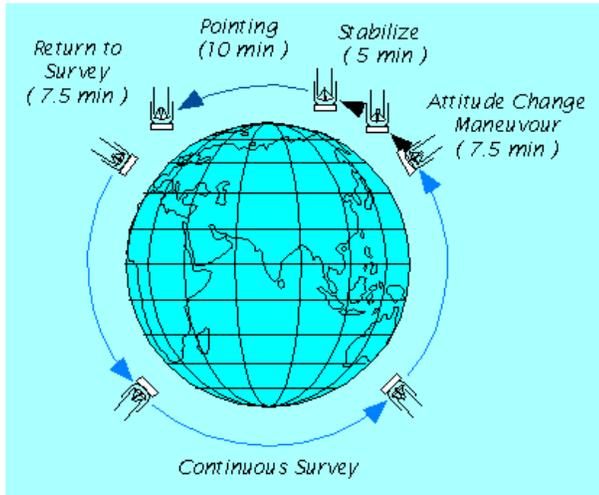}}
\end{center}
\caption{The orbit and the attitude control sequence of ASTRO-F.
	The basic observation mode is the continuous survey, 
	but also the pointing observation for less than 10 minutes
	can be made.}
\label{fig:orbit}
\end{figure}

ASTRO-F is to be launched into the sun synchronous polar orbit with
the altitude of 750~km. The trajectory of ASTRO-F is alway on 
the twilight zone. This orbit is similar to that of IRAS, and
is suitable for all sky survey observations with severe constraints
of avoidance angles
toward the sun and toward the earth, as in the case
of infrared satellites with cooled telescopes.
FIS performs the all-sky survey in the continuous survey mode as
is shown in Figure~8. 
We will concentrate on this mode of observations
at least in the first half year of the mission.

For any observations with IRC or spectroscopic observations
with FIS, we need pointing observation.
In the case of the pointing mode of ASTRO-F, 
the line of sight of the telescope can be fixed 
to a specific direction on the sky
for about 10 minutes (Figure~8).    
The nominal attitude is that the line of sight of the telescope
is perpendicular to the direction toward the sun.
We can change this angle only by $\pm 1\degr$.
Hence the time slot for specific objects to be observed 
in the pointing mode is strongly restricted.
The maximum number of pointing observations we can
make in one orbit is three.

\subsection{Capability}

\begin{figure}[tbp]
\begin{center}
\resizebox{8.5cm}{!}{\includegraphics*[73,192][753,498]{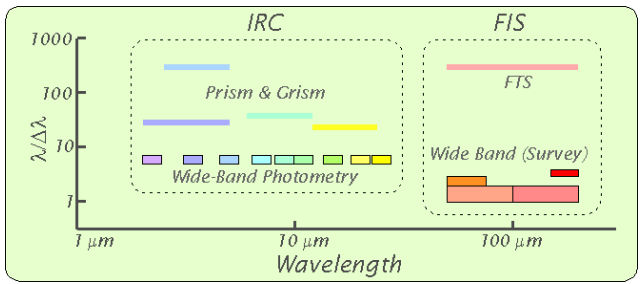}}
\end{center}
\caption{Schematic view of the wavelength coverage and the spectroscopic
	capability of IRC and FIS onboard ASTRO-F.}
\label{fig:capability}
\end{figure}

\begin{figure}[htb]
\begin{center}
\epsfig{file=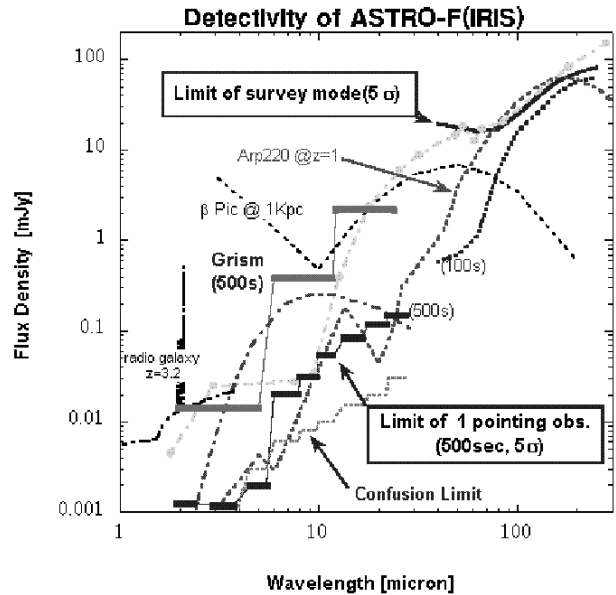, width=8.5cm}
\end{center}
\caption{Sensitivity of ASTRO-F for point sources. The FIS
	sensitivity during the survey mode
	at the wavelength longer than 100~$\mu$m is limited by
	the confusion noise, and the sensitivity cannot be improved
	even with longer integration time.}
\label{fig:sensitivity}
\end{figure}

Figure~9 shows a schematic 
view of the wavelength coverage and the spectroscopic
capability of IRC and FIS onboard ASTRO-F. 
ASTRO-F covers wide spectral range both in the photometric
mode and in the spectroscopic mode with moderate resolution.

Figure~10 shows the sensitivity of 
ASTRO-F for point sources. 
The FIS
sensitivity during the survey mode
at the wavelength longer than 100~$\mu$m is limited by
the confusion noise, and the sensitivity cannot be improved
even with longer integration time.
The sensitivity of the all sky survey of ASTRO-F/FIS is better
than that of IRAS by more than an order of magnitude. 
Hence the results of the ASTRO-F/FIS will serve
as a legacy for the astronomical community for a long time
and will be ideal inputs for observation planning of FIRST.

IRC also achieved almost confusion-limited sensitivity
over a wide range of spectral region. The big advantage 
of IRC is that it has a wide field of view ($10' \times 10'$).
The database of IRC observations will be 
very important to determine
the spectral energy distribution
of many sources
to be observed by FIRST at longer wavelengths.



\appendix

\section{Comparison with SIRTF}

NASA's infrared mission SIRTF (Gallagher \& Simmons 2000) 
is to be launched in 2002. The telescope size of SIRTF is 85~cm, which
is slightly larger than that of ASTRO-F, and the spectral 
coverage of SIRTF is similar to that of ASTRO-F. Hence
ASTRO-F and SIRTF have very similar specifications in many points.
SIRTF has better sensitivity than ASTRO-F especially in far-infrared
due to its slightly larger telescope size and finer spatial
sampling. On the other hand, the FOV of the focal plane instruments
on ASTRO-F is wider that those on SIRTF. Hence ASTRO-F
is more suitable for efficient observations of wide areas. Especially,
the all sky survey at far-infrared is a unique program of ASTRO-F. 

\section{Collaboration with ESA}

We have started the discussion with ESA on possible
collaboration concerning the data analysis activity and 
supports of tracking stations for ASTRO-F.
The goal of this collaboration is to enable
quick release of ASTRO-F survey catalogues.
We are also discussing the possibility to open
some fraction of pointing observations
for European community.

\end{document}